\documentclass[reprint,pra,superscriptaddress,amsmath,amssymb,aps,longbibliography,]{revtex4-1}

\usepackage{graphicx}% Include figure files
\usepackage{dcolumn}% Align table columns on decimal point
\usepackage{bm}% bold math
\usepackage{color}
\usepackage{blindtext}
\usepackage{appendix}
\usepackage{lipsum}
\usepackage{xr}
\usepackage{xcolor}
\usepackage{siunitx}
\usepackage{multirow}
\usepackage[colorlinks]{hyperref}% add hypertext capabilities
\hypersetup{%
	plainpages=true,
	breaklinks=true,% not default in dvips mode, so we must specify
	hypertexnames=false,%not ideal, but needed when pagenums duplicate (`i' vs. `1')
	pageanchor=true,
	colorlinks=true,
	linkcolor={blue},
	citecolor={blue},
	urlcolor={blue},
	anchorcolor={black}
}

\newcommand{\bt}{\textbf}

\begin{document}

\title{Low-Loss, High-Coherence Airbridge Interconnects Fabricated by Single-Step Lithography}

\author{J.~B.~Fu} 
\affiliation{College of Computer Science and Technology, National University of Defense Technology, Changsha 410073, China}
\author{B.~Ren} 
\affiliation{College of Computer Science and Technology, National University of Defense Technology, Changsha 410073, China}
\author{J.~D.~Ouyang} 
\affiliation{College of Computer Science and Technology, National University of Defense Technology, Changsha 410073, China}
\author{C.~Li} 
\affiliation{College of Computer Science and Technology, National University of Defense Technology, Changsha 410073, China}
\author{K.~C.~Q.~Zhu} 
\affiliation{College of Computer Science and Technology, National University of Defense Technology, Changsha 410073, China}
\author{Y.~G.~Che}
\affiliation{College of Computer Science and Technology, National University of Defense Technology, Changsha 410073, China}
\author{X.~Fu}
\affiliation{College of Computer Science and Technology, National University of Defense Technology, Changsha 410073, China}
\author{S.~C.~Xue} 
\affiliation{College of Computer Science and Technology, National University of Defense Technology, Changsha 410073, China}
\author{Z.~H.~Yang} 
\email{zhaohuayang@quanta.org.cn}
\affiliation{College of Computer Science and Technology, National University of Defense Technology, Changsha 410073, China}
\author{M.~T.~Deng} 
\email{mtdeng@nudt.edu.cn}
\affiliation{College of Computer Science and Technology, National University of Defense Technology, Changsha 410073, China}
\author{J.~J.~Wu}
\affiliation{College of Computer Science and Technology, National University of Defense Technology, Changsha 410073, China}

% \date{\today}

\begin{abstract}
Airbridges are essential for creating high-performance, low-parasitic interconnects in integrated circuits and quantum devices. Conventional multi-step fabrication methods hinder miniaturization and introduce process-related defects. We report a simplified process for fabricating nanoscale airbridges using only a single electron-beam lithography step. By optimizing a multilayer resist stack with a triple-exposure-dose scheme and a thermal reflow step, we achieve smooth, suspended metallic bridges with sub-200-nm features that exhibit robust mechanical stability. Fabricated within a gradiometric SQUID design for superconducting transmon qubits, these airbridges introduce no measurable additional loss in the relaxation time $T_1$, while enabling a 2.5‑fold enhancement of the dephasing time $T_2^*$. This efficient method offers a practical route toward integrating high‑performance three‑dimensional interconnects in advanced quantum and nano‑electronic devices.
\end{abstract}

\maketitle
\section*{Introduction}
 The continuous drive for miniaturization, increased functionality, and superior performance in modern integrated circuits, photonic integrated circuits, and micro-electromechanical systems necessitates the development of complex three-dimensional architectures. A critical challenge is creating high-performance, low-parasitic interconnections that can span over underlying components or other interconnect layers without introducing detrimental electrical, optical, or mechanical interference. Airbridge structures are an indispensable enabling technology for this purpose~\cite{Sherwin1994, Villeneuve1995, BORZENKO2004210, Girgis2006}. In quantum information processing, airbridges are particularly important for realizing low-loss microwave resonators, essential coupling elements between qubits, and effective grounding strategies that suppress spurious electromagnetic modes, thereby directly enhancing qubit coherence and gate fidelity~\cite{chen2014fabrication, Sun2022Fabrication, Stavenga2023Lower, Huang2025Fabrication, Bolgar2025Air}.

Conventional airbridge fabrication typically relies on multi-layer resist stacks and multiple lithography steps to define suspended metallic structures after selective etching of a sacrificial layer~\cite{ZHANG2007194, Dunsworth2018A, Tao2024fabrication, bu2025tantalum, bruckmoser2025}. While functional, these methods require complex alignment across exposures and risk introducing organic residues or process-induced defects that can degrade quantum coherence~\cite{Siddiqi2021, Mahuli2025}, thereby hindering further miniaturization and high-density integration. Alternative single-step methods, such as gradient exposure~\cite{Janzen2022Aluminum} or two-photon lithography~\cite{Huang2025Fabrication}, can form arched bridge profiles but face challenges in scaling to sub-micron dimensions.

Here, we present a method for fabricating airbridges using only a single electron-beam lithography (EBL) step. By optimizing a multilayer resist composition and applying a triple-exposure-dose scheme followed by a thermal reflow, we achieve an undercut profile that yields smooth, suspended metallic bridges after development and lift-off. These single-lithography airbridges (SL-airbridges) feature sub-200-nm dimensions, a well-defined arch shape, and strong resilience to low-power ultrasonic treatment. This streamlined process can be directly integrated into quantum circuit fabrication without introducing additional decoherence sources.

%%%%%%%%%%%%%%%%%%%%%% FIG. 1 %%%%%%%%%%%%%%%%%%%%%%%%
\begin{figure}[t]
\includegraphics[width=1\linewidth]{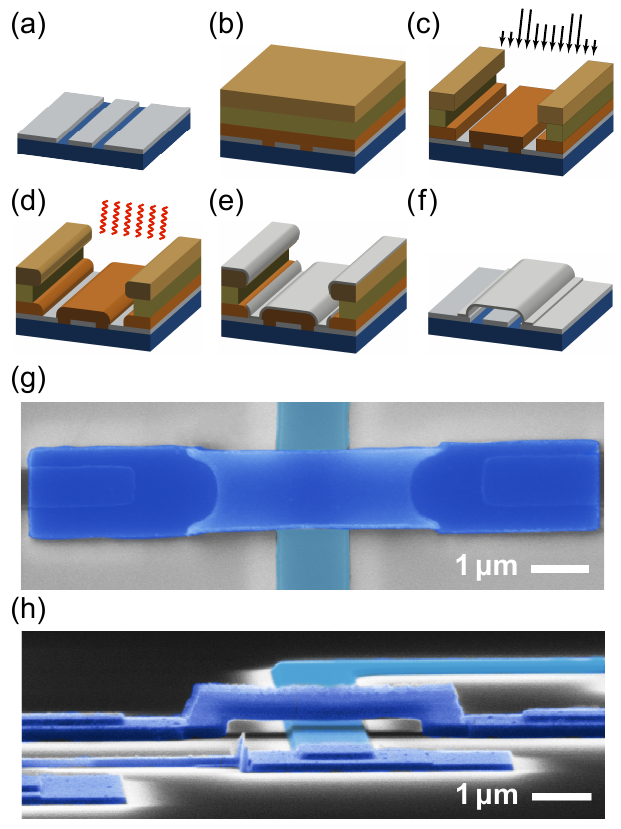}
\caption{\bt{Schematic of the single-lithography airbridge (SL-airbridge) fabrication process. } \bt{a}, Deposition, patterning, and etching of a 100-nm-thick Al base wire.
	\bt{b}, Spin-coating of a multilayer electron-beam resist stack.
	\bt{c}, Patterning via EBL using three distinct exposure doses.
	\bt{d}, Thermal reflow of the developed resist stack on a hotplate.
	\bt{e}, Ion milling of the base metal contact areas followed by electron-beam evaporation of the bridge metal (150~nm Al).
	\bt{f}, Lift-off and ultrasonic cleaning.
	\bt{g}, Top-view scanning electron microscopy (SEM) image of a fabricated nanoscale airbridge.
	\bt{h}, Side-view SEM image of an airbridge taken at an 85\textdegree~stage tilt. }
\label{Fig1}
\end{figure}
%%%%%%%%%%%%%%%%%%%%%%%%%%%%%%%%%%%%%%%%%%%%%%%%%%%

\section*{Experiment setup}
The SL-airbridge fabrication sequence is illustrated in Fig.~\ref{Fig1}. We start from introducing the carefully designed electron-beam resist stack. First, a 300-nm-thick layer of poly(methyl methacrylate) (PMMA, 950k molecular weight) is spin-coated on a substrate and baked at 160~\textcelsius~for 10~minutes. This bottom layer defines the electrical contact region between the airbridge and the underlying circuit. Next, a 400-nm-thick layer of methyl methacrylate (MMA EL9) is spin-coated and baked under the same conditions. Finally, a 400-nm-thick blended resist layer, composed of PMMA and MMA in a 2:3 ratio, is spin-coated and baked. The blended top layer offers higher sensitivity than pure PMMA for forming the bridge profile while maintaining pattern stability during the subsequent thermal reflow.

Patterning is performed using standard EBL. A critical aspect is the application of three calibrated exposure doses. The highest dose (exceeding 1000~$\mu$C/cm$^2$) fully exposes the entire resist stack to define the airbridge pedestal. An intermediate dose (300--500~$\mu$C/cm$^2$) exposes the MMA and PMMA/MMA blend layers to pattern the suspended bridge profile. The lowest dose (below 150~$\mu$C/cm$^2$) is designed to create a controlled undercut within the middle MMA layer, which is crucial for successful lift-off after reflow.

After exposure, the sample is developed in a 1:3 mixture of methyl isobutyl ketone and isopropyl alcohol (MIBK:IPA) for 45~s. A thermal reflow step is then performed at 125~\textcelsius, the glass transition temperature of PMMA, for a precisely controlled duration of 45~s. This step transforms the developed profile into a smooth, arched shape essential for continuous metal deposition.
%%%%%%%%%%%%%%%%%%%%%% FIG. 2 %%%%%%%%%%%%%%%%%%%%%%%%
\begin{figure}[ht]
\includegraphics[width=\linewidth]{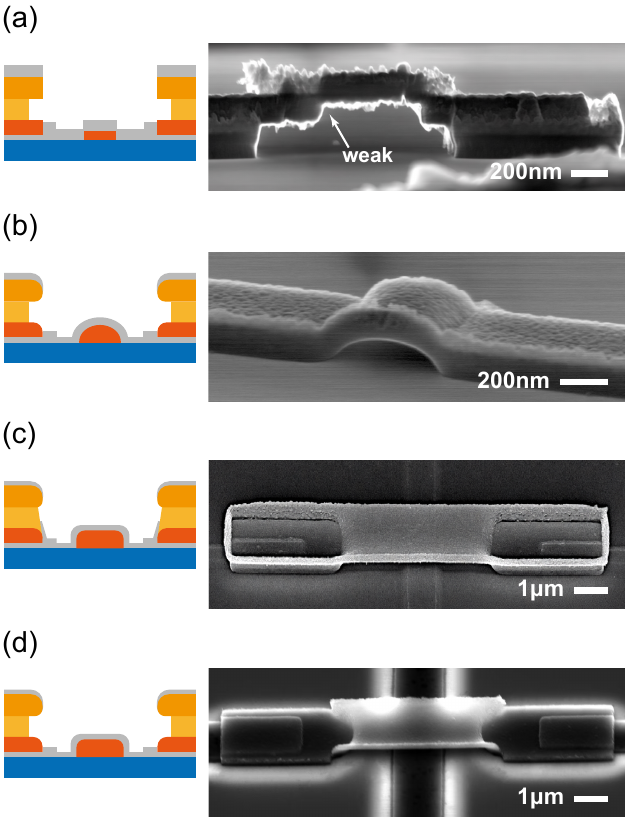}
\caption{\bt{Control experiments of reflow and triple-dose exposure scheme.} Comparison of SL-airbridges fabricated \bt{a} without and \bt{b} with the thermal reflow step and comparison of SL-airbridges fabricated \bt{c} with a double-dose exposure scheme and \bt{d} with the optimized triple-dose exposure scheme.}
\label{Fig2}
\end{figure}
%%%%%%%%%%%%%%%%%%%%%%%%%%%%%%%%%%%%%%%%%%%%%%%%%%%

%%%%%%%%%%%%%%%%%%%%%% FIG. 3 %%%%%%%%%%%%%%%%%%%%%%%%
\begin{figure}[ht]
	\includegraphics[width=\linewidth]{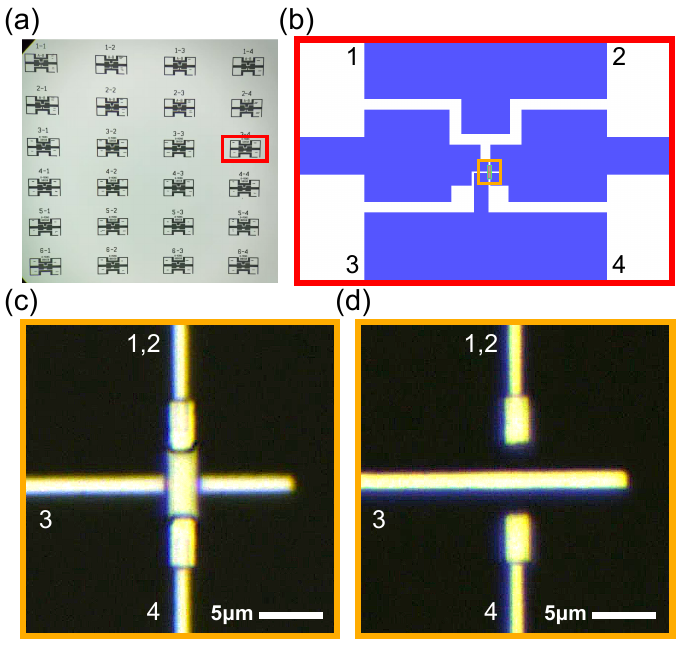}
	\caption{\bt{Ultrasonic treatment tests of SL-airbridges.} \bt{a}, Optical images of a subset of the 60 test samples.
		\bt{b}, Layout of a single test sample with four contact pads.
		\bt{c}, Intact SL-airbridge after low-power ultrasonic treatment for 20~s.
		\bt{d}, Damaged SL-airbridge after high-power ultrasonic treatment for 20~s. }
	\label{Fig3}
\end{figure}
%%%%%%%%%%%%%%%%%%%%%% FIG. 3 %%%%%%%%%%%%%%%%%%%%%%%%
Metallization begins with argon ion milling to clean the contact areas and ensure low-resistance interfaces, followed by electron-beam evaporation of a 150-nm-thick aluminum film at normal incidence. Lift-off is carried out in an 80~\textcelsius~N-methyl-2-pyrrolidone (NMP) bath. Low-power ultrasonic agitation is applied to remove edge residues and improve the bridge smoothness.

Figures~\ref{Fig1}(g) and (h) show representative SEM images of a completed SL-airbridge, which is 2~$\mu$m wide and approximately 200~nm high. The smooth edges indicate a high-quality lift-off process.

A previously reported single-step lithography method using a triple-layer PMMA/MMA stack employed only two exposure doses and omitted thermal reflow~\cite{Jin2021}. The resulting airbridges exhibited rough edges, which could increase surface loss and electromagnetic radiation, thereby degrading quantum coherence. Furthermore, such structures could not tolerate ultrasonic cleaning, limiting their use in complex fabrication sequences.

To highlight the importance of thermal reflow and the triple-dose strategy, we performed control experiments, shown in Fig.~\ref{Fig2}. Without reflow, the deposited metal layers show rough edges and discontinuous connections [Fig.~\ref{Fig2}(a)]. With reflow, the SL-airbridge maintains a smooth profile even at sub-micron scales [Fig.~\ref{Fig2}(b)]. However, reflow alone tends to reduce the undercut, complicating lift-off and leading to raised metal walls, as seen in Fig.~\ref{Fig2}(c). The triple-dose exposure scheme resolves this conflict by using a low dose to preserve a controlled undercut in the MMA layer after reflow. Airbridges fabricated with a double-dose scheme exhibit unstable, raised edges due to the resist slope [Fig.~\ref{Fig2}(c)], whereas the triple-dose scheme yields smooth, well-defined edges [Fig.~\ref{Fig2}(d)]. 

%%%%%%%%%%%%%%%%%%%%%% FIG. 4 %%%%%%%%%%%%%%%%%%%%%%%%
\begin{figure*}[ht]
	\includegraphics[width=0.95\linewidth]{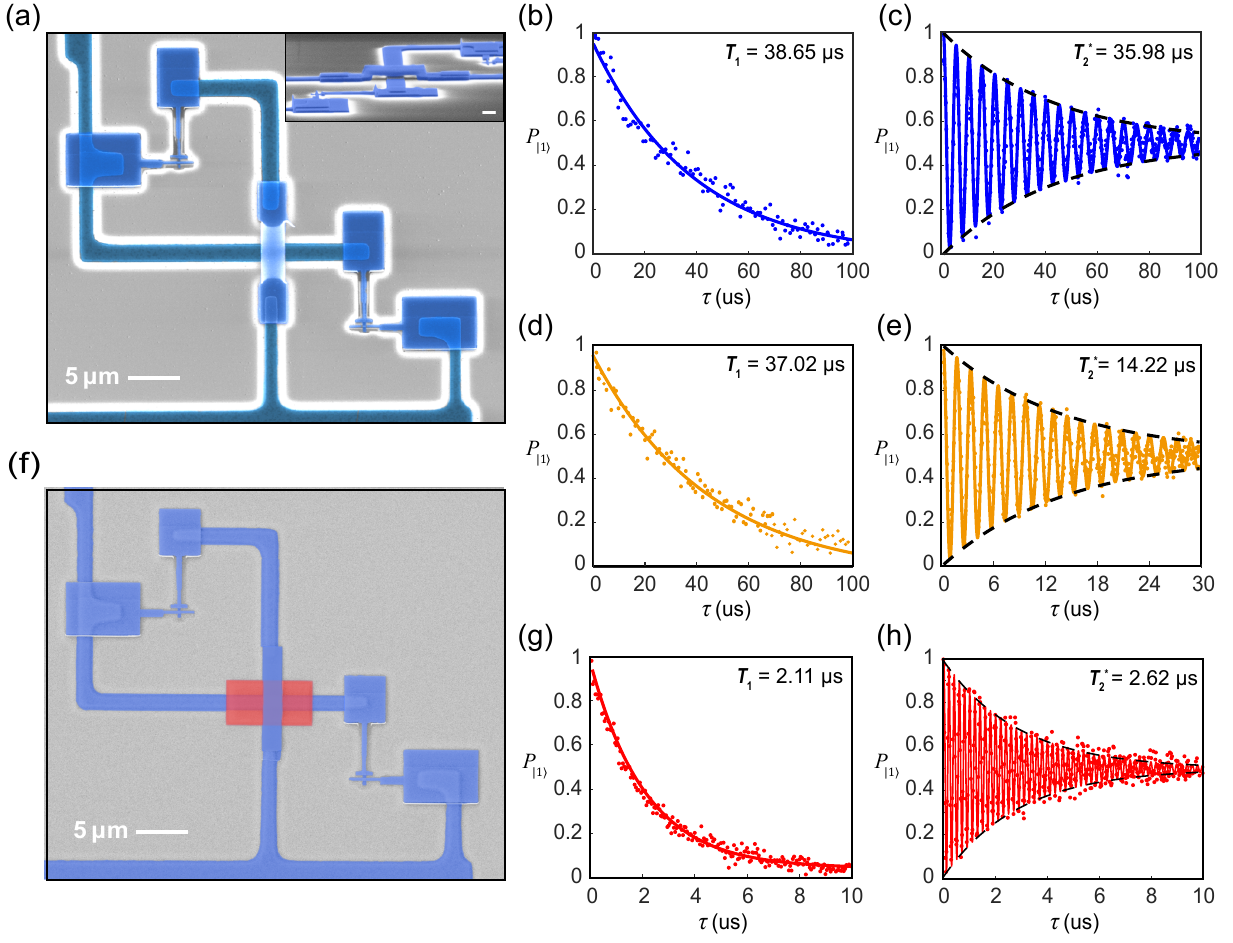}
	\caption{\bt{Enhanced qubit coherence with SL-airbridge interconnects.}
		\bt{a}, False-color SEM image of an 8-mon qubit with an SL-airbridge (blue) connecting the two loops of the gradiometric SQUID. Inset shows a side-view SEM image at a 75\textdegree~tilt.
		\bt{b}, Highest measured $T_1$ for qubit $Q_{1}$ (8-mon with SL-airbridge).
		\bt{c}, Highest measured $T_2^*$ for $Q_{1}$.
		\bt{d}, Highest measured $T_1$ for qubit $Q_{4}$ (X-mon).
		\bt{e}, Highest measured $T_2^*$ for $Q_{4}$.
		\bt{f}, False-color SEM image of an 8-mon qubit with a dielectric (SiO$_2$, red) crossover.
		\bt{g}, Highest measured $T_1$ for qubit $Q_{6}$ (8-mon with dielectric crossover).
		\bt{h}, Highest measured $T_2^*$ for $Q_{6}$.}
	\label{Fig4}
\end{figure*}
%%%%%%%%%%%%%%%%%%%%%%%%%%%%%%%%%%%%%%%%%%%%%%%%%%%

The robustness of the SL-airbridge against ultrasonic processing, a common step in superconducting device fabrication, was systematically evaluated. We prepared 60 test samples, each containing four electrode pads for electrical characterization [Figs.~\ref{Fig3}(a,b)].
Pads 1 and 2 are electrically shorted through the airbridge. Conductivity between pad 1 (or 2) and pad 4 confirms the airbridge's electrical integrity, while resistance between pad 1 (or 2) and pad 3 checks for shorts indicative of collapse or damage. All SL-airbridges on the sample are 5 $\mu$m in length and 2 $\mu$m in width. Samples were treated in an ultrasonic cleaner (Branson 3800) for 20~s in low-power mode. All 60 samples survived intact [Fig.~\ref{Fig3}(c)]. When treated for 20~s in high-power mode, approximately 30\% of the airbridges were damaged [Fig.~\ref{Fig3}(d)], demonstrating a practical operating limit.

\section*{Characterization of qubit coherence}

To demonstrate the utility of SL-airbridges in quantum devices, we implemented them in a novel transmon qubit design featuring a doubly connected gradiometric superconducting quantum interference device (SQUID), termed the ``8-mon''~\cite{fu2026flux}. This design aims to enhance coherence by canceling quasi-static flux noise while preserving frequency tunability. As shown in the false-colored SEM image in Fig.~\ref{Fig4}(a), an SL-airbridge connects the two loops of the gradiometric SQUID.

\begin{table}[b]
	\vspace{-20pt}
	\caption{
		Summary of single-qubit parameters from coherence measurements. Values are averages over more than 10 measurements. $\omega^{\mathrm{idle}}$ is the idle frequency, and $\omega^{\mathrm{R}}$ is the readout resonator frequency.
	}
	\label{table1}
	\begin{ruledtabular}
		\begin{tabular}{ccccc}
			\multirow{2}{*}{Qubit}&$\omega^{\mathrm{idle}}/2\pi$ &$\omega^{\mathrm{R}}/2\pi$&$\bar{T}_{1}$
			&$\bar{T}_{2}^{*}$\\
			& $(\mathrm{GHz})$ & $(\mathrm{GHz})$ & $(\mu\mathrm{s})$ & $(\mu\mathrm{s})$ \\ \hline
			$Q_{1}$\footnotemark[1]& 5.028 & 6.430 & 33.70 & 32.40 \\
			$Q_{2}$\footnotemark[1]& 4.473 & 6.286 & 33.57 & 27.73 \\
			$Q_{3}$\footnotemark[1]& 4.426 & 6.384 & 26.57 & 26.51 \\
			$Q_{4}$\footnotemark[2]& 5.397 & 6.615 & 27.91 & 13.05 \\
			$Q_{5}$\footnotemark[2]& 4.954 & 6.572 & 22.82 & 11.48 \\ 
			$Q_{6}$\footnotemark[3]& 4.448 & 6.271 & 1.96 & 2.34 \\
			$Q_{7}$\footnotemark[3]& 4.911 & 6.222 & 1.88 & 1.30 \\
			$Q_{8}$\footnotemark[3]& 4.993 & 6.415 & 1.53 & 1.63 \\
		\end{tabular}
	\end{ruledtabular}
	\footnotetext[1]{8-mon qubit with an SL-airbridge crossover.}
	\footnotetext[2]{X-mon qubit.}
	\footnotetext[3]{8-mon qubit with a dielectric (SiO$_2$) crossover.}
\end{table}

We co-fabricated 8-mon qubit and conventional X-mon qubit on the same chip and measured their coherence properties under identical conditions. Results for selected qubits are summarized in Table~\ref{table1}, with all data averaged over more than 10 measurement repetitions. The $T_1$ times of the 8-mons are comparable to those of the X-mons. Qubit $Q_{1}$ (8-mon) achieved a maximum $T_1$ of 38.65~$\mu$s [Fig.~\ref{Fig4}(b)], indicating that the SL-airbridge introduces no measurable additional relaxation loss. The same qubit reached a maximum $T_2^*$ of 35.98~$\mu$s [Fig.~\ref{Fig4}(c)], approximately 2.5 times that of the X-mon ($Q_{4}$, $T_2^* = 14.22~\mu$s) [Fig.~\ref{Fig4}(e)]. This confirms the efficacy of the gradiometric design in mitigating dephasing noise. The coherence times of our X-mons are consistent with recent reports~\cite{wu2021strong, shi2023quantum, xiang2023simulating, Wu_2025}.

To underscore the low-loss nature of the airbridge interconnect, we fabricated a control sample where the crossover in the 8-mon was formed using a dielectric insulating layer (SiO$_2$) deposited via electron-beam evaporation at 95~\textcelsius [Fig.~\ref{Fig4}(f)]. The introduction of this amorphous dielectric material severely degraded coherence, reducing both $T_1$ and $T_2^*$ to approximately 2~$\mu$s [Figs.~\ref{Fig4}(g,h) and Table~\ref{table1}], highlighting the superior performance of the airbridge-based approach.

\section*{Summary}
In summary, we have developed a streamlined process for fabricating airbridge interconnects requiring only single-step lithography. The method combines an optimized multilayer resist stack, a triple-exposure-dose scheme, and a thermal reflow step to reliably produce smooth, suspended metallic bridges with sub-200-nm features and lengths exceeding 5~$\mu$m. Validated in superconducting quantum devices, the SL-airbridge exhibits low loss, high mechanical stability, and full compatibility with standard fabrication workflows. It enables greater design freedom for complex quantum circuit geometries. This approach offers a practical and accessible route for integrating high-performance three-dimensional interconnects in next-generation quantum circuits, photonic devices, and nano-electronics, where minimizing parasitic interactions is paramount for advancing performance and integration density.

\newpage
\bibliography{bibfile}

@article{Sherwin1994,
    author = {Sherwin, M. E. and Simmons, J. A. and Eiles, T. E. and Harff, N. E. and Klem, J. F.},
    title = {Parallel quantum point contacts fabricated with independently biased gates and a submicrometer airbridge post},
    journal = {Applied Physics Letters},
    volume = {65},
    number = {18},
    pages = {2326-2328},
    year = {1994},
    month = {10},
    doi = {10.1063/1.112731},
}

@article{Villeneuve1995,
    author = {Villeneuve, Pierre R. and Fan, Shanhui and Joannopoulos, J. D. and Lim, Kuo‐Yi and Petrich, G. S. and Kolodziejski, L. A. and Reif, Rafael},
    title = {Air‐bridge microcavities},
    journal = {Applied Physics Letters},
    volume = {67},
    number = {2},
    pages = {167-169},
    year = {1995},
    month = {07},
    doi = {10.1063/1.114655},
}

@article{BORZENKO2004210,
title = {Metallic air-bridges fabricated by multiple acceleration voltage electron beam lithography},
journal = {Microelectronic Engineering},
volume = {75},
number = {2},
pages = {210-215},
year = {2004},
doi = {https://doi.org/10.1016/j.mee.2004.05.005},
author = {T Borzenko and C Gould and G Schmidt and L.W Molenkamp}
}

@article{Girgis2006,
    author = {Girgis, E. and Liu, J. and Benkhedar, M. L.},
    title = {Fabrication of metallic air bridges using multiple-dose electron beam lithography},
    journal = {Applied Physics Letters},
    volume = {88},
    number = {20},
    pages = {202103},
    year = {2006},
    month = {05},
    doi = {10.1063/1.2204833},
}

@article{Siddiqi2021,
author = {Siddiqi, Irfan},
doi = {10.1038/s41578-021-00370-4},
journal = {Nature Reviews Materials},
month = {sep},
number = {10},
pages = {875--891},
publisher = {Nature Research},
title = {{Engineering high-coherence superconducting qubits}},
volume = {6},
year = {2021}
}

@article{Mahuli2025,
archivePrefix = {arXiv},
arxivId = {2506.17474},
author = {Mahuli, Neha and Minguzzi, Joaquin and Gao, Jiansong and Resnick, Rachel and Diez, Sandra and Cosmic, R. and Marcaud, Guillaume and Hunt, Matthew and Swenson, Loren and Rose, Jefferson and Painter, Oskar and Jarrige, Ignace},
doi = {10.1021/acsnano.5c14022},
eprint = {2506.17474},
journal = {ACS Nano},
month = {dec},
number = {48},
pages = {41136--41146},
pmid = {41263414},
title = {{Improving the Lifetime of Aluminum-Based Superconducting Qubits through Atomic Layer Etching and Deposition}},
volume = {19},
year = {2025}
}

@article{chen2014fabrication,
    author = {Chen, Zijun and Megrant, A. and Kelly, J. and Barends, R. and Bochmann, J. and Chen, Yu and Chiaro, B. and Dunsworth, A. and Jeffrey, E. and Mutus, J. Y. and O'Malley, P. J. J. and Neill, C. and Roushan, P. and Sank, D. and Vainsencher, A. and Wenner, J. and White, T. C. and Cleland, A. N. and Martinis, John M.},
    title = {Fabrication and characterization of aluminum airbridges for superconducting microwave circuits},
    journal = {Applied Physics Letters},
    volume = {104},
    number = {5},
    pages = {052602},
    year = {2014},
    month = {02},
    doi = {10.1063/1.4863745},
}

@article{Sun2022Fabrication,
    author = {Sun, Yuting and Ding, Jiayu and Xia, Xiaoyu and Wang, Xiaohan and Xu, Jianwen and Song, Shuqing and Lan, Dong and Zhao, Jie and Yu, Yang},
    title = {Fabrication of airbridges with gradient exposure},
    journal = {Applied Physics Letters},
    volume = {121},
    number = {7},
    pages = {074001},
    year = {2022},
    month = {08},
    doi = {10.1063/5.0102555},
}

@article{Janzen2022Aluminum,
    author = {Janzen, N. and Kononenko, M. and Ren, S. and Lupascu, A.},
    title = {Aluminum air bridges for superconducting quantum devices realized using a single-step electron-beam lithography process},
    journal = {Applied Physics Letters},
    volume = {121},
    number = {9},
    pages = {094001},
    year = {2022},
    month = {08},
    doi = {10.1063/5.0103165},
}

@article{Stavenga2023Lower,
    author = {Stavenga, T. and Khan, S. A. and Liu, Y. and Krogstrup, P. and DiCarlo, L.},
    title = {Lower-temperature fabrication of airbridges by grayscale lithography to increase yield of nanowire transmons in circuit QED quantum processors},
    journal = {Applied Physics Letters},
    volume = {123},
    number = {2},
    pages = {024004},
    year = {2023},
    month = {07},
    doi = {10.1063/5.0146814},
}

@article{Huang2025Fabrication,
    author = {Huang, Yi-Hsiang and Wang, Haozhi and Shen, Zhuo and Thomas, Austin and Richardson, C. J. K. and Palmer, B. S.},
    title = {Fabrication of metal air bridges for superconducting circuits using two-photon lithography},
    journal = {Applied Physics Letters},
    volume = {127},
    number = {4},
    pages = {044002},
    year = {2025},
    month = {07},
    doi = {10.1063/5.0271788},
}

@article{ZHANG2007194,
title = {Fabrication of nanoscale metallic air-bridges by introducing a SiO2 sacrificial layer},
journal = {Materials Science in Semiconductor Processing},
volume = {10},
number = {4},
pages = {194-199},
year = {2007},
doi = {https://doi.org/10.1016/j.mssp.2007.11.004},
author = {Yang Zhang and Jian Liu and Yan Li and Fuhua Yang}
}

@article{Dunsworth2018A,
    author = {Dunsworth, A. and Barends, R. and Chen, Yu and Chen, Zijun and Chiaro, B. and Fowler, A. and Foxen, B. and Jeffrey, E. and Kelly, J. and Klimov, P. V. and Lucero, E. and Mutus, J. Y. and Neeley, M. and Neill, C. and Quintana, C. and Roushan, P. and Sank, D. and Vainsencher, A. and Wenner, J. and White, T. C. and Neven, H. and Martinis, John M. and Megrant, A.},
    title = {A method for building low loss multi-layer wiring for superconducting microwave devices},
    journal = {Applied Physics Letters},
    volume = {112},
    number = {6},
    pages = {063502},
    year = {2018},
    month = {02},
    doi = {10.1063/1.5014033},
}

@article{Tao2024fabrication,
    author = {Tao, Hao-Ran and Zhang, Chi and Du, Lei and Yang, Xin-Xin and Guo, Liang-Liang and Chen, Yong and Zhang, Hai-Feng and Jia, Zhi-Long and Kong, Wei-Cheng and Duan, Peng and Guo, Guo-Ping},
    title = {Fabrication and characterization of low loss niobium airbridges for superconducting quantum circuits},
    journal = {Applied Physics Letters},
    volume = {125},
    number = {3},
    pages = {034001},
    year = {2024},
    month = {07},
    doi = {10.1063/5.0216711},
}

@article{bu2025tantalum,
  title={Tantalum airbridges for scalable superconducting quantum processors},
  author={Bu, Kunliang and Huai, Sainan and Zhang, Zhenxing and Li, Dengfeng and Li, Yuan and Hu, Jingjing and Yang, Xiaopei and Dai, Maochun and Cai, Tianqi and Zheng, Yi-Cong and others},
  journal={npj Quantum Information},
  year={2025},
  month={Jan},
  day={29},
  volume={11},
  number={1},
  pages={17},
  publisher={Nature Publishing Group UK London},
  doi={10.1038/s41534-025-00972-8},
}

@article{bruckmoser2025,
  title={Niobium Air Bridges as a Low-Loss Component for Superconducting Quantum Hardware}, 
  author={N. Bruckmoser and L. Koch and I. Tsitsilin and M. Grammer and D. Bunch and L. Richard and J. Schirk and F. Wallner and J. Feigl and C. M. F. Schneider and S. Geprägs and V. P. Bader and M. Althammer and L. Södergren and S. Filipp},
  eprint={2503.12076},
  archivePrefix={arXiv},
  year={2025},
}

@article{Bolgar2025Air,
    author = {Bolgar, Aleksey N. and Kalacheva, Daria A. and Lubsanov, Viktor B. and Dmitriev, Aleksei Yu. and Alekseeva, Evgenia S. and Korostylev, Evgeny V. and Astafiev, Oleg V.},
    title = {Highly stable aluminum air-bridges with stiffeners},
    journal = {Journal of Applied Physics},
    volume = {137},
    number = {15},
    pages = {154401},
    year = {2025},
    month = {04},
    doi = {10.1063/5.0260833},
}

@article{fu2026flux,
  title={Flux-noise-resilient transmon qubit via a doubly-connected gradiometric design}, 
  author={J. B. Fu and Da-Wei Wang and B. Ren and Z. H. Yang and S. Hu and G. Y. Huang and S. H. Cao and D. D. Liu and X. F. Zhang and X. Fu and S. C. Xue and Y. G. Che and Yu-xi Liu and M. T. Deng and J. J. Wu},
  year={2026},
  eprint={2601.02137},
  archivePrefix={arXiv},
}

@article{wu2021strong,
  title = {Strong Quantum Computational Advantage Using a Superconducting Quantum Processor},
  author = {Wu, Yulin and Bao, Wan-Su and Cao, Sirui and Chen, Fusheng and Chen, Ming-Cheng and Chen, Xiawei and Chung, Tung-Hsun and Deng, Hui and Du, Yajie and Fan, Daojin and Gong, Ming and et.al},
  journal = {Physical Review Letters},
  volume = {127},
  issue = {18},
  pages = {180501},
  numpages = {7},
  year = {2021},
  month = {Oct},
  doi = {10.1103/PhysRevLett.127.180501},
}

@article{shi2023quantum,
  title = {Quantum Simulation of Topological Zero Modes on a 41-Qubit Superconducting Processor},
  author = {Shi, Yun-Hao and Liu, Yu and Zhang, Yu-Ran and Xiang, Zhongcheng and Huang, Kaixuan and Liu, Tao and Wang, Yong-Yi and Zhang, Jia-Chi and Deng, Cheng-Lin and Liang, Gui-Han and Mei, Zheng-Yang and Li, Hao and Li, Tian-Ming and Ma, Wei-Guo and Liu, Hao-Tian and Chen, Chi-Tong and Liu, Tong and Tian, Ye and Song, Xiaohui and Zhao, S. P. and Xu, Kai and Zheng, Dongning and Nori, Franco and Fan, Heng},
  journal = {Physical Review Letters},
  volume = {131},
  issue = {8},
  pages = {080401},
  numpages = {6},
  year = {2023},
  month = {Aug},
  publisher = {American Physical Society},
  doi = {10.1103/PhysRevLett.131.080401},
}

@article{xiang2023simulating,
  title={Simulating Chern insulators on a superconducting quantum processor},
  author={Xiang, Zhong-Cheng and Huang, Kaixuan and Zhang, Yu-Ran and Liu, Tao and Shi, Yun-Hao and Deng, Cheng-Lin and Liu, Tong and Li, Hao and Liang, Gui-Han and Mei, Zheng-Yang and others},
  journal={Nature Communication},
  year={2023},
  month={Sep},
  day={05},
  volume={14},
  number={1},
  pages={5433},
  doi={10.1038/s41467-023-41230-9},
}

@article{Jin2021,
    author = {Jin, Y. and Moreno, M. and T. Vianez, P. M. and Tan, W. K. and Griffiths, J. P. and Farrer, I. and Ritchie, D. A. and Ford, C. J. B.},
    title = {Microscopic metallic air-bridge arrays for connecting quantum devices},
    journal = {Applied Physics Letters},
    volume = {118},
    number = {16},
    pages = {162108},
    year = {2021},
    month = {04},
    doi = {10.1063/5.0045557},
}

@article{Wu_2025,
doi = {10.1088/0256-307X/42/4/040501},
year = {2025},
month = {apr},
publisher = {Chinese Physical Society and IOP Publishing Ltd},
volume = {42},
number = {4},
pages = {040501},
author = {Wu, Zhi-Hao and Lei, Ling-Xiao and Zhang, Xin-Fang and Xue, Shi-Chuan and Hu, Shun and Li, Cong and Fu, Xiang and Chen, Ping-Xing and Lu, Kai and Deng, Ming-Tang and Wu, Jun-Jie},
title = {Telegraph Flux Noise Induced Beating Ramsey Fringe in Transmon Qubits},
journal = {Chinese Physics Letters}
}
\bibliographystyle{naturemag}

\section*{Acknowledgments}
We thank Yang Yang, Xiao-Feng Yi, Ding-Dong Liu, Peng Luo, and Kang-Ding Zhao for technical support in device fabrication and measurements. This work was supported by the National Key R\&D Program of China (Grant No. 2024YFB4504000) and the Aid Program for Science and Technology Innovative Research Team in Higher Educational Institutions of Hunan Province. Device fabrication was partially performed at the Synergetic Extreme Condition User Facility (SECUF).

\section*{Author Contributions} 
M.-T.D. conceived the central concepts and coordinated the experimental work. Device fabrication was carried out by J.-B.F. and B.R., while J.-B.F. and M.-T.D. performed the measurements. D.-W.W. and Y.-X.L. conducted the theoretical modeling and numerical simulation simulations. S.H., D.-D.L., J.-B.F., and Z.-H.Y. developed the software under the supervision of X.F. All authors participated in interpreting the results and in the preparation of the manuscript.

\clearpage

% \captionsetup[figure]{name=Fig. S}

% %%%%%%%%%%%%%%%%%%%%%% FIG. S1 %%%%%%%%%%%%%%%%%%%%%%%%
% \begin{figure}[t]
% \includegraphics[width=0.75\linewidth]{FigS1-3.pdf}
% \caption{\bt{Experimental conditions.} \bt{a}, Chip-A conditions,including magnetic shielding box}
% \label{figS1}
% \end{figure}
% %%%%%%%%%%%%%%%%%%%%%%%%%%%%%%%%%%%%%%%%%%%%%%%%%%%

\end{document}